\documentclass[superscriptaddress,aps,preprintnumbers,amsmath,showpacs,amssymb,prd,nofootinbib,reprint]{revtex4-1}
\usepackage{bm,color} 
\usepackage{amssymb,amsfonts,slashed,amsthm,amsmath,graphicx, soul}
\usepackage{epsfig}
\usepackage{ulem}
\usepackage{float}

\begin{document}

\renewcommand{\figurename}{Fig.}
\renewcommand{\tablename}{Table.}
\newcommand{\Slash}[1]{{\ooalign{\hfil#1\hfil\crcr\raise.167ex\hbox{/}}}}
\newcommand{\bra}[1]{ \langle {#1} | }
\newcommand{\ket}[1]{ | {#1} \rangle }
\newcommand{\beq}{\begin{equation}}  \newcommand{\eeq}{\end{equation}}
\newcommand{\bef}{\begin{figure}}  \newcommand{\eef}{\end{figure}}
\newcommand{\bec}{\begin{center}}  \newcommand{\eec}{\end{center}}
\newcommand{\non}{\nonumber}  \newcommand{\eqn}[1]{\begin{equation} {#1}\end{equation}}
\newcommand{\laq}[1]{\label{eq:#1}}  
\newcommand{\dd}[1]{{d \o d{#1}}}
\newcommand{\Eq}[1]{Eq.~(\ref{eq:#1})}
\newcommand{\Eqs}[1]{Eqs.~(\ref{eq:#1})}
\newcommand{\eq}[1]{(\ref{eq:#1})}
\newcommand{\Sec}[1]{Sec.\ref{chap:#1}}
\newcommand{\ab}[1]{\left|{#1}\right|}
\newcommand{\vev}[1]{ \left\langle {#1} \right\rangle }
\newcommand{\bs}[1]{ {\boldsymbol {#1}} }
\newcommand{\lac}[1]{\label{chap:#1}}
\newcommand{\SU}[1]{{\rm SU{#1} } }
\newcommand{\SO}[1]{{\rm SO{#1}} }
\def\({\left(}
\def\){\right)}
\def\dt{{d \o dt}}
\def\diag{\mathop{\rm diag}\nolimits}
\def\Spin{\mathop{\rm Spin}}
\def\O{\mathcal{O}}
\def\U{\mathop{\rm U}}
\def\Sp{\mathop{\rm Sp}}
\def\SL{\mathop{\rm SL}}
\def\tr{\mathop{\rm tr}}
\def\ebq{\end{equation} \begin{equation}}
\newcommand{\OR}{~{\rm or}~}
\newcommand{\AND}{~{\rm and}~}
\newcommand{\EV}{ {\rm \, eV} }
\newcommand{\KEV}{ {\rm \, keV} }
\newcommand{\MEV}{ {\rm \, MeV} }
\newcommand{\GEV}{ {\rm \, GeV} }
\newcommand{\TEV}{ {\rm \, TeV} }
\def\o{\over}
\def\a{\alpha}
\def\b{\beta}
\def\c{\varepsilon}
\def\d{\delta}
\def\e{\epsilon}
\def\f{\phi}
\def\g{\gamma}
\def\h{\theta}
\def\k{\kappa}
\def\l{\lambda}
\def\m{\mu}
\def\n{\nu}
\def\p{\psi}
\def\q{\partial}
\def\r{\rho}
\def\s{\sigma}
\def\t{\tau}
\def\u{\upsilon}
\def\w{\omega}
\def\x{\xi}
\def\y{\eta}
\def\z{\zeta}
\def\D{\Delta}
\def\G{\Gamma}
\def\H{\Theta}
\def\L{\Lambda}
\def\F{\Phi}
\def\P{\Psi}
\def\S{\Sigma}
\def\me{\mathrm e}
\def\ol{\overline}
\def\tl{\tilde}
\def\*{\dagger}


\preprint{TU-1250}

\title{
Stimulated Emission of Dark Matter via Thermal Scattering: \\ Novel Limits for Freeze-In and eV Cold Dark Matter}

\author{
Kodai Sakurai
}
\affiliation{Department of Physics, Tohoku University, 
Sendai, Miyagi 980-8578, Japan} 
\affiliation{ Institute of Theoretical Physics, Faculty of Physics, University of Warsaw, ul. Pasteura 5, PL-02-093 Warsaw, Poland}
\author{
Wen Yin
}
\affiliation{Department of Physics, Tokyo Metropolitan University, 
Minami-Osawa, Hachioji-shi, Tokyo 192-0397 Japan}

\begin{abstract}
Recently, one of the present authors noticed a stimulated emission process of bosonic dark matter via the two-body decay of a mother particle in a thermal plasma similar to the operation principle of a laser in \cite{Yin:2023jjj}.  In this paper, we show that in a $2 \to 2$  process, including a bosonic final particle (e.g., an axion or dark photon), the stimulated emission occurs as well due to a small angle scattering of the thermal mother particles and thus the phenomenon is more universal. Two important conclusions follow: (1) Care must be taken when studying the freeze-in production of a bosonic dark matter, as the abundance and momentum distribution of dark matter can differ significantly due to this effect. (2) eV-mass-range bosonic dark matter is special and theoretically well-motivated because models for freeze-in or other thermal production of dark matter include the parameter region of cold eV dark matter. 
We also study the dark matter mass effect for the stimulated emission. 

\end{abstract}

\maketitle
\flushbottom

\vspace{1cm}

\section{Introduction}

Dark matter plays a crucial role in the evolution of the Universe around and after matter-radiation equality. However, the particle properties of dark matter, such as its mass, interactions, and spin, as well as its production mechanisms in the early Universe, remain unknown. For instance, relatively heavy dark matter has been widely searched for such as in dark matter detection experiments~\cite{PandaX-4T:2021bab,LZ:2022lsv,XENON:2023cxc,Fermi-LAT:2015att,HESS:2022ygk} and collider experiments~\cite{ATLAS:2024fdw,CMS:2024zqs}.
Recently, light dark matter candidates, such as axions or hidden photons, get more attention. 

Several attempts have been made to produce light dark matter in simple setups that account for quantum statistical effects, such as Bose enhancement or the Pauli exclusion principle. In Refs.~\cite{Moroi:2020has, Moroi:2020bkq, Nakayama:2021avl, Choi:2023jxw}, the authors studied the decay of the inflaton or dark Higgs to produce light dark matter by taking into account quantum statistical effects in the Boltzmann equation and found that the resulting dark matter can be much lighter than previously thought, consistent with constraints from small-scale structures. In Refs.~\cite{Batell:2024hzo, Cho:2024ehb}, the authors have shown that the robust stability of dark matter decaying into lighter fermions can be achieved due to the Pauli exclusion principle arising from the Fermi sea of the daughter particles, which may consist of the cosmic neutrino background. One of the authors, WY, showed in Ref.~\cite{Yin:2023jjj} that the thermal production of dark matter, i.e., the production of the dark matter by particles in thermal equilibrium, is significantly impacted by quantum statistical effects by solving the unintegrated Boltzmann equation in the context of the $1 \to 2$ decay process.

In particular, it has been shown that eV-mass-range dark matter can be produced due to a `stimulated emission' of dark matter in a much lower momentum mode compared to the cosmic temperature~\cite{Yin:2023jjj}. The mechanism for this production is similar to that of a laser. The reason for the eV mass range is the same as the conventional `hot dark matter paradigm': the dark matter number density is around the one of the thermal plasma, and the matter radiation equality is around eV cosmic temperature. In the stimulated emission, the dark matter distribution reaches a {\it quasi-equilibrium}, with the number density close to the thermal plasa. However, dark matter produced in this way can suppress the free-streaming length, making eV-range dark matter consistent with the structure formation. To ensure that the {\it quasi-equilibrium} is not disturbed over cosmological timescales, both final states in the $1 \to 2$ process should be weakly coupled.

On the other hand, a well-known thermal production mechanism is the freeze-in mechanism~\cite{Moroi:1993mb,Dodelson:1993je, Hall:2009bx,Lebedev:2019ton,Ballesteros:2020adh,Sakurai:2021ipp,Bhatia:2023yux,Dror:2023fyd,Ai:2023qnr, Abe:2020ldj}, in which dark matter is so weakly coulped that it never reaches thermal equilibrium with the thermal bath in the early Universe. Instead, its number density is thermally produced much smaller than the thermal plasma density, and thus the resulting dark matter mass can be much heavier than eV. To estimate the abundance precisely, one needs to solve the Boltzmann equation. Typically, some approximations of the equation and ansatz for the particle distribution function are used. As a result, the aforementioned quantum statistical effects are neglected or effectively neglected depending on the ansatz. Indeed, in \cite{Yin:2023jjj}, it was shown in a simple $1 \to 2$ process that the freeze-in production could be affected by quantum statistical effects. In particular, when the weakly-coupled bosonic dark matter is pair produced, which is the leading contribution with the dark matter being stabilized by some symmetry, the afformentioned condition for the quasi-equilibrium is automatically met.

It is questioned whether the stimulated emission from the thermal plasma is a more universal phenomenon or just for the $1\to 2$ process which has a restricted phase space. 
 This motivates us to study the $2 \to 2$ process. Moreover, gauge singlet light dark matter can only be pair produced from two or more gauge charged mother particles.

In this paper, we show that the stimulated emission of the bosonic dark matter production can occur as well in the $2\to 2$ process due to small angle scatterings of the thermal mother particles, if the production is not too UV momentum dominated. We derive various conditions for the stimulated emission to occur in the system. 
Avoiding the stimulated emission, since otherwise the keV-GeV dark matter would be significantly overproduced, we derive a novel bound on the freeze-in mechanism. In addition, we show that eV-range bosonic dark matter is still special and theoretically well-motivated in a more universal sense than in \cite{Yin:2023jjj}. Several new ingredients of the stiumlated emission of dark matter are found which can also apply to $1\to 2$ case: mass effect of the dark matter, thermal mass effect of the mother particles.

This paper is organized as follows. In the next section, we discuss the setup we will consider. In Sec.\,\ref{sec:num}, the numerical result will be given. The behavior will be explained analytically and various formulas will be derived in Sec.\,\ref{sec:Results}. 
The formulas are used to constrain the Freeze-in scenario by avoiding the stimulated emissions in~\Sec{limit}. 
In \Sec{app}, we will study the mass effect of the dark matter on the stimulated emission and application to produce the eV cold dark matter. The last section is devoted to the conclusions.

\section{Setup}\label{sec:setup}

\subsection{Class of models of interest}
In this paper, we solve the unintegrated Boltzmann equation without neglecting Bose enhancement or Pauli blocking factors to study dark matter thermal production via the simple setup
\beq 
 \chi_1 \chi_2 \to \phi_1 \phi_2,\label{eq:process}
\eeq 
in which one of the $\phi_{1,2}$ particles is a bosonic dark matter particle. 
$\chi_{1,2}$ are the mother particles with interactions so fast that they always form a thermal distribution. 

For generality, we consider the (spin-)averaged amplitude squared in the form
\beq \laq{ampsq}
|{\cal M}|^2= c_n (m_{\chi_1} m_{\chi_2})^{-n} s^{n}\;,
\eeq
where $c_{n}$, $s$, and $m_{\chi_{1,2}}$ are the prefactor of the squared amplitude, the Mandelstam variable, and the masses of $\chi_{1,2}$, respectively. $n$ is a real parameter, which may or may not be an integer. For simplicity of notation we include the symmetry factor in $|{\cal M}|^2$, e.g., for $\f_1=\f_2$ case, $|{\cal M}|^2$ is the half of that derived from the Feynmann rule. 
Therefore, the cross section is of the form
\beq
\sigma v_{\rm rel}= c_n {g_{\f_1} g_{\f_2}}\frac{(m_{\chi_1} m_{\chi_2})^{-n}}{32\pi E_{\chi_1} E_{\chi_2}} s^{n} \laq{crosssect}
\eeq
 and $g_{\f_{1,2}}$ is the internal degrees of freedom of $\f_{1,2}$. We have neglect the masses of the $\phi_{1,2}$ particles here. Namely we consider the production of the dark matter with cosmic temperature much higher than the dark matter mass. 
In addition, we will consider $\phi_1 = \phi_2 = \phi$ for the motivation discussed above. 
For instance, we may consider the Higgs portal interaction $\chi_{1,2}^2 \phi^2$~\cite{Silveira:1985rk,Burgess:2000yq}, and then $\chi_{1,2}$ can be understood as some Higgs fields, which corresponds to the $n=0$ case. 
In particular, $\phi$ can be an axion (more precisely, see, e.g., \cite{Sakurai:2021ipp,Sakurai:2022cki,Haghighat:2022qyh}). 

We will come back to the effect of the dark matter mass and the case of $\phi_1 \neq \phi_2$ in \Sec{app}. 

\subsection{Boltzmann equation}
We describe analytical formulae for the Boltzmann equations for the two-boson production process \eq{process}. In the expanding universe, it can be written as~(see, e.g.,~\cite{Kolb:1990vq})
\begin{align} \label{eq:Bolzeq}
\frac{\partial f_i(p_i,t)}{\partial t} - p_i H \frac{\partial f_i(p_i,t)}{\partial p_i} = C^i(p_i,t)
\end{align}
where $f_i(p_i)$ is the distribution function (the momentum) for a particle $i = \chi_1, \chi_2, \phi$, and $H$ is the Hubble parameter defined by the scale factor, $H = \dot{a}/a$. 
The left-hand side corresponds to the Liouville operator, and the right-hand side corresponds to the collision term. 
Furthermore, we assume that the distribution function has rotational invariance. 
Thus, $f_i$ is given by a function of the absolute value of the momentum $|\vec{p}_i| \equiv p_i$.

The collision term for the boson $\phi$ from \eq{process} is given by
\begin{align} \label{eq:colisont}
&C^{\phi}(p_{\phi_1},t) = \frac{1}{2 E_{\phi_1} g_{\phi}} \sum \int d\Pi_{\phi_2} \, d\Pi_{\chi_1} \, d\Pi_{\chi_2} \notag \\
&\quad \times (2\pi)^4 \delta^{(4)}(p_{\chi_1} + p_{\chi_2} - p_{\phi_1} - p_{\phi_2}) |{\cal M}|^2 
 \times S\end{align}
where the phase space integral is defined by $d\Pi_i = \frac{d^3 p_i}{ 2 E_i (2\pi)^{3}}$. We introduced the index of $1,2$ for the $\f$ momentum for clarity. 
The summation symbol denotes the sum over the internal degrees of freedom for $\chi_{1,2}$ and two $\phi$s.
The function $S$ is given in terms of the distribution functions as
\begin{align}
S \equiv
f_{\chi_1}(p_{\chi_1}) f_{\chi_2}(p_{\chi_2})
\left(1 + f_{\phi}(p_{\phi_1}) \right) \left(1 + f_{\phi}(p_{\phi_2}) \right) \notag \\
- f_{\phi}(p_{\phi_1}) f_{\phi}(p_{\phi_2})
\left(1 \pm f_{\chi_1}(p_{\chi_1}) \right) \left(1 \pm f_{\chi_2}(p_{\chi_2}) \right) \;.
\end{align}
In the case where $\chi_1, \chi_2$ are bosons (fermions), $+$ ($-$) is taken, thereby involving the effect of Bose enhancement (Pauli exclusion) in the Boltzmann equation.

{In addition to this collision term, we also need to include terms for reactions, e.g., $\phi \chi_1 \to \phi \chi_2$, which may contribute to processes of the same order. They are model-dependent, given that we parametrize the amplitude of \Eq{process} by the generic form \eq{ampsq}, and since these terms should not change our conclusions, as we will see in \Sec{scat}, we do not consider those contributions in this paper. However it is interestinng to perform a comprehensive study in a concrete model like the Higgs portal, which will be discussed elsewhere.} 

\subsection{Phase space integration of the collision term}
We here discuss the phase space integration of the collision term.
We choose $E_{\phi_2}$, $E_{\chi_1}$, and $\cos{\theta_{\phi_1 \phi_2}}$ as integration variables in the phase space integrations, where $\theta_{\phi_1 \phi_2}$ is defined as the angle between $\vec{p}_{\phi_1}$ and $\vec{p}_{\phi_2}$. 
We can express the Lorentz-invariant phase space integrations in the collision term in terms of these kinematical variables as
\begin{align}
&d\Pi_{\phi_2}=\frac{1}{(2\pi)^2}\frac{{p}_{\phi_2}}{2}dE_{\phi_2}\cos{\theta_{\phi_1\phi_2}}\;, \notag \\
&d\Pi_{\chi_1}d\Pi_{\chi_2}(2\pi)^4\delta^{(4)} (p_{\chi_1}+p_{\chi_2}-p_{\phi_1}-p_{\phi_2}) \notag \\
&\quad\quad=\frac{1}{8\pi}\frac{1}{p_{\phi_1+\phi_2}}dE_{\chi_1}\;,
\end{align}
with $p_{\phi_1 + \phi_2}$ being 
\begin{align}
p_{\phi_1 + \phi_2} = \left( p_{\phi_1}^2 + p_{\phi_2}^2 + 2 p_{\phi_1} p_{\phi_2} \cos{\theta_{\phi_1 \phi_2}} \right)^{1/2} \;.
\end{align}

Correcting these expressions, we can reduce the collision term as
\begin{align}
C^{\phi} &= \frac{{g_\f} {g_{\chi_1} g_{\chi_2}}}{2 E_{\phi_1}} \int_{p_{\phi_2}^{\rm min}}^{p_{\phi_2}^{\rm max}} dp_{\phi_2}
\int_{s^-}^{s^+} ds  
\int_{E_{\chi_1}^{\rm min}}^{E_{\chi_1}^{\rm max}} dE_{\chi_1} \notag \\
&\quad \times \frac{1}{16 (2\pi)^3} \frac{p_{\phi_2}}{E_{\phi_2} \sqrt{E_{\phi_1+\phi_2}^2 - s}} 
\, |{\cal M}|^2 \notag \\
&\quad \times S\left(f_{\chi_1}(p_{\chi_1}), f_{\chi_2}(p_{\chi_2}), f_{\phi}(p_{\phi_1}), f_{\phi}(p_{\phi_2})\right) \;,
\end{align}
with 
\[
E_{\phi_1+\phi_2} = E_{\phi_1} + E_{\phi_2}.
\]
{Here, we replace $\cos\theta_{\phi_1\phi_2}$ by}
the Mandelstam variable $s$, which is given by 
\[
s = (E_{\phi_1} + E_{\phi_2})^2 - |\vec{p}_{\phi_1} + \vec{p}_{\phi_2}|^2.
\]
The integration limits of $s$ are calculated as
\begin{align}
s^\pm = 2 m_\phi^2 + 2 E_{\phi_1} E_{\phi_2} \pm 2 p_{\phi_1} p_{\phi_2} \;. 
\end{align}
The upper (lower) limit of $E_{\chi_1}$, labeled by $E_{\chi_1}^{\rm max}$ ($E_{\chi_1}^{\rm min}$), is determined from the energy and momentum conservation laws, i.e., $E_{\chi_1} + E_{\chi_2} = E_{\phi_1} + E_{\phi_2}$ and $\vec{p}_{\chi_1} + \vec{p}_{\chi_2} = \vec{p}_{\phi_1} + \vec{p}_{\phi_2}$.
These conservation laws yield the following relation 
\begin{align} 
\label{eq:eqEchi1}
(E_{\phi_1} + E_{\phi_2} - E_{\chi_1})^2 - m_{\chi_2}^2 &= p_{\phi_1+\phi_2}^2 + p_{\chi_1}^2 \notag \\
&\quad - 2 p_{\phi_1+\phi_2} p_{\chi_1} \cos{\theta_{\chi_1}},
\end{align} 
with $\theta_{\chi_1}$ being the angle between $\vec{p}_{\chi_1}$ and $\vec{p}_{\phi_1} + \vec{p}_{\phi_2}$. 
The energy $E_{\chi_1}$ reaches its lower or upper limit at $|\cos{\theta_{\chi_1}}| = 1$. 
Solving Eq.~\eqref{eq:eqEchi1}, one can obtain
\begin{align} 
E_{\chi_1}^{\rm max/min} &= \frac{1}{2 s} \Bigg[
E_{\phi_1+\phi_2} (s + m_{\chi_1}^2 - m_{\chi_2}^2) \notag \\
&\quad \pm \sqrt{ (E_{\phi_1+\phi_2}^2 - s) \lambda(s, m_{\chi_1}^2, m_{\chi_2}^2) }
\Bigg]
\end{align}
with $\lambda$ being 
\[
\lambda(x, y, z) = (x - y - z)^2 - 4 y z.
\]
The remaining unknown parameter $E_{\chi_2}$ is given by 
\[
E_{\chi_2} = \sqrt{ p_{\phi_1}^2 - m_{\phi}^2 } + \sqrt{ p_{\phi_2}^2 - m_{\phi}^2 } - E_{\chi_1} \;. 
\]

\section{Numerical simulations}\label{sec:num}

\begin{figure*}[htbp] \includegraphics[scale=0.72]{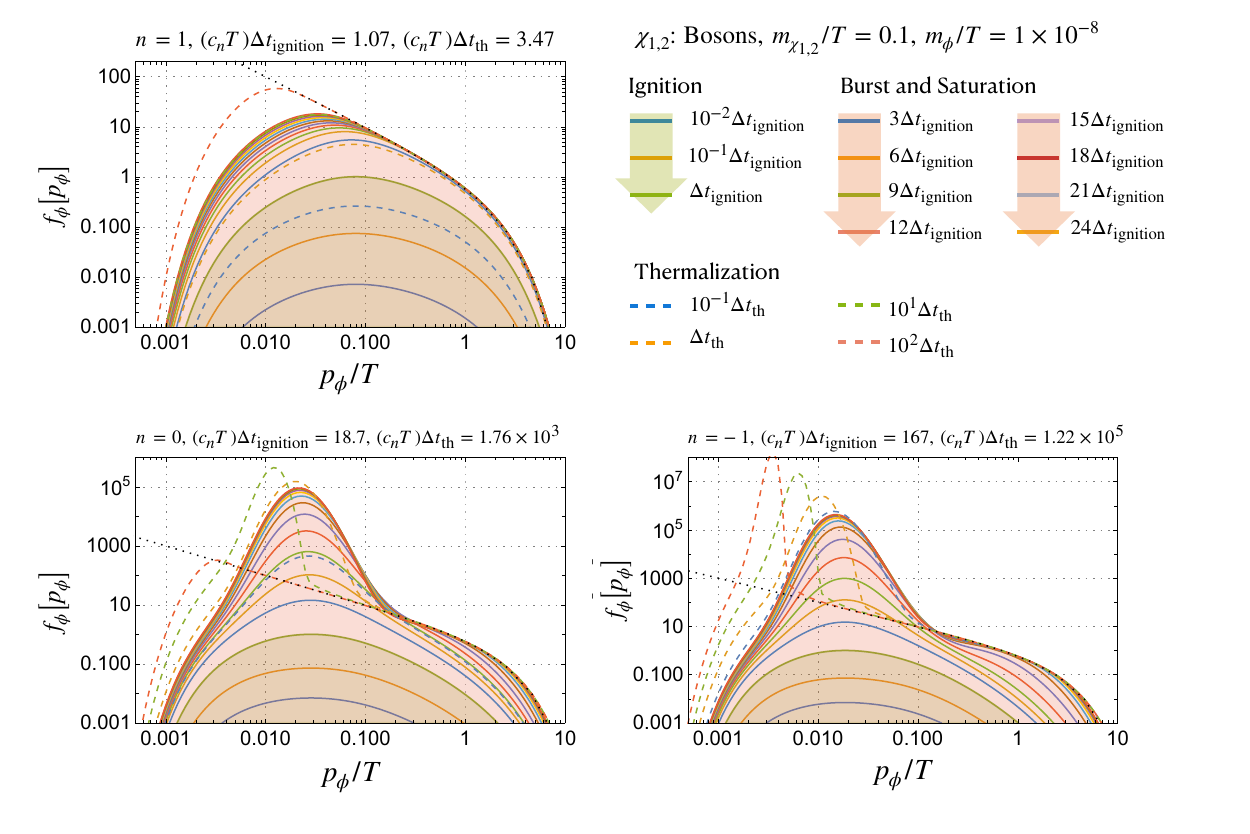} \caption{Distribution function $f_{\phi}$ as a function of {$p_{\phi}/T$}, where $\chi_{1,2}$ are identified as bosons. For the phase space integration in the collision term, we set $m_{\chi_1} = m_{\chi_2} = 1 \times 10^{-1} {T},\ {m_\phi = 1 \times 10^{-8} {T}}$. We take $g_{\chi_1,\chi_2,\f} = 1$. In addition, we set the range of $p_{\phi}$ as $[10^{-6}, 10^{2}]$, and the step size as $200$. The black dotted line shows the distribution function at thermal equilibrium, ${f_{\phi}^{\rm eq} = 1/(\exp(\sqrt{p_\phi^2 + m_\phi^2}/T) - 1)}$. } \label{fig:fphiBoson} \end{figure*}

\begin{figure*}[htbp] \includegraphics[scale=0.72]{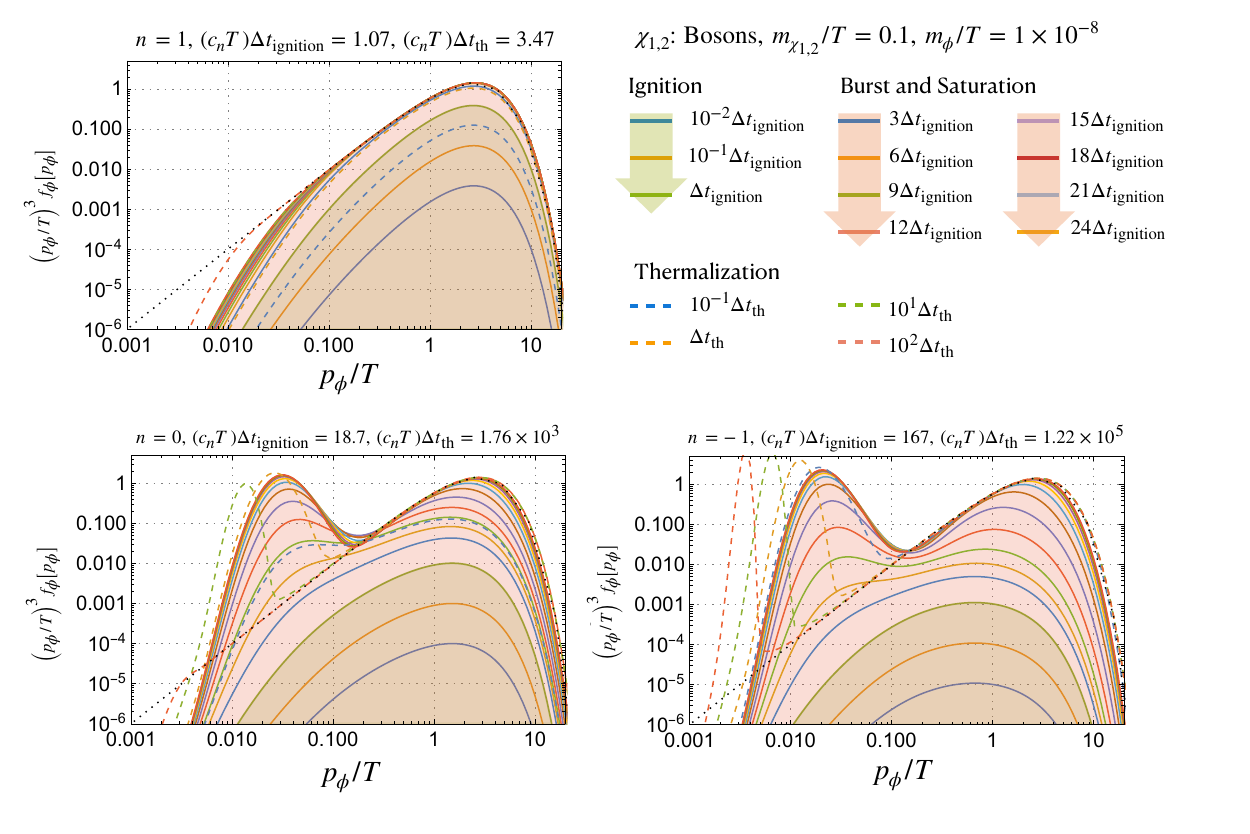} \caption{The same figure as Fig.~\ref{fig:fphiBoson} but for {$(p_{\phi}/T)^3 f_\phi$}. } \label{fig:p3fphiBoson} \end{figure*}

\begin{figure*}[t] \includegraphics[scale=0.72]{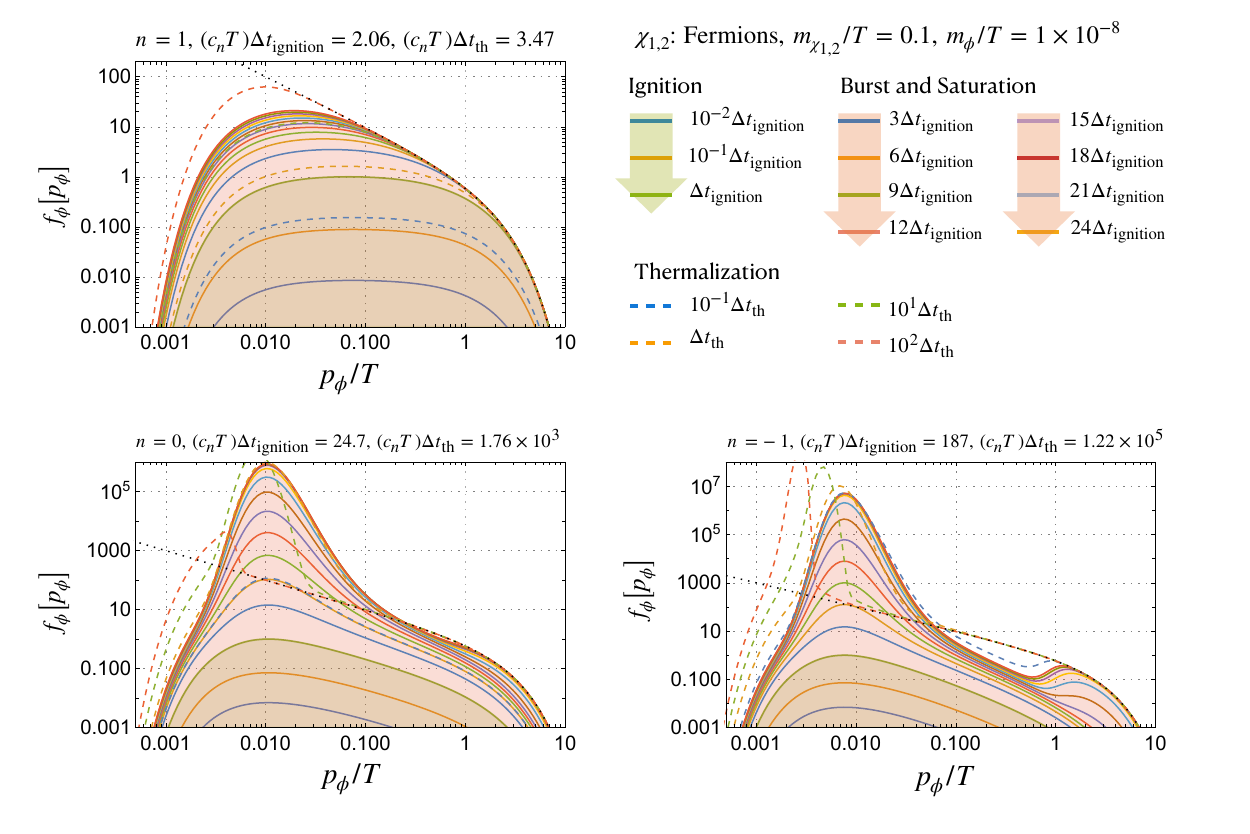} \caption{Distribution function $f_{\phi}$ as a function of {$p_{\phi}/T$}, where $\chi_{1,2}$ are identified as fermions. For the phase space integration in the collision term, we set $m_{\chi_1} = m_{\chi_2} = 1 \times 10^{-1} {T},\ {m_\phi = 1 \times 10^{-8} {T}}$, and {$c_{n} = 1$}. In addition, we set the range of $p_{\phi}$ as $[10^{-6}, 10^{2}]$, and the step size as $200$. The black dotted line shows the distribution function at thermal equilibrium, ${f_{\phi}^{\rm eq} = 1/(\exp(\sqrt{p_\phi^2 + m_\phi^2}/T) - 1)}$. We take $g_{\chi_1,\chi_2} = 2, g_\f = 1$. } \label{fig:fphiFermion}
 \end{figure*}

\begin{figure*}[tbph] \includegraphics[scale=0.72]{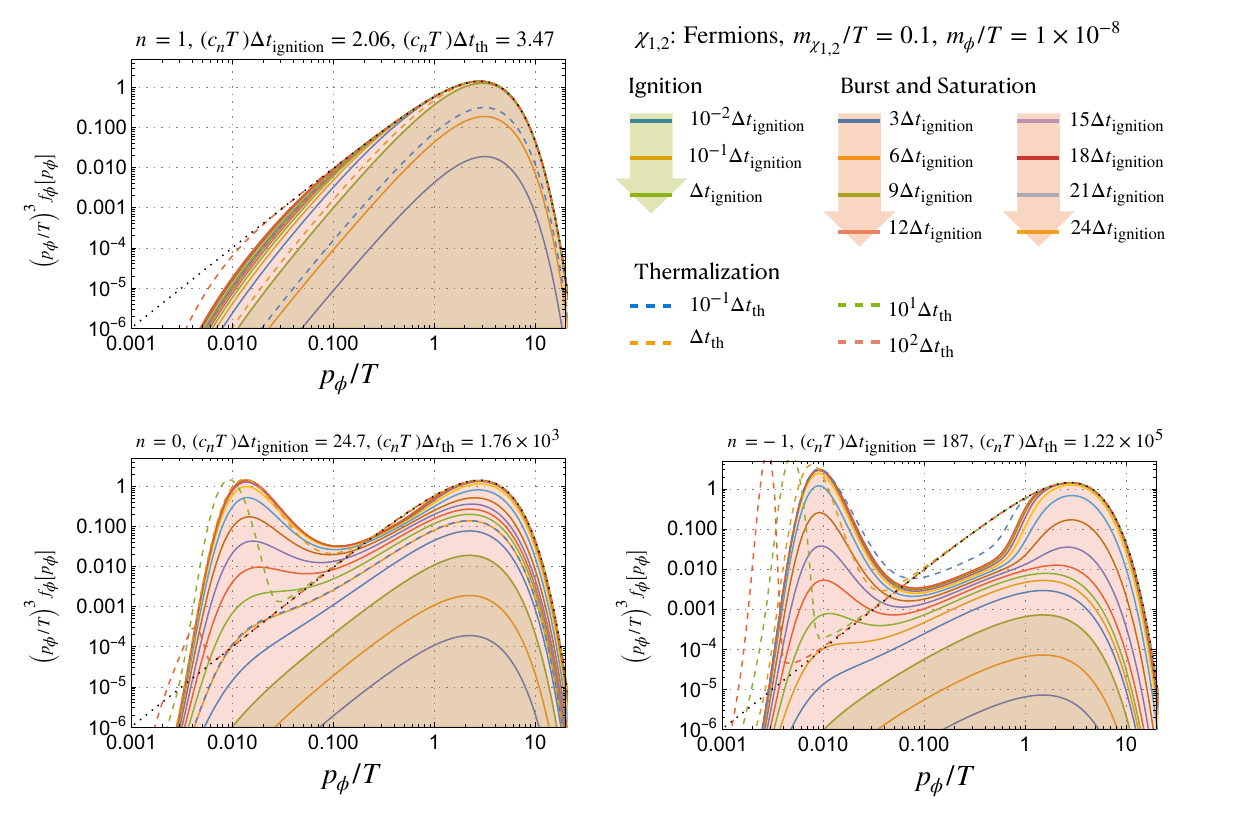} 
\caption{The same figure as Fig.~\ref{fig:fphiFermion}
 but for {$(p_{\phi}/T)^3 f_\phi$}.  \label{fig:p3fphiFermion}} \end{figure*}

In this section, we numerically solve the Boltzmann equation for $\phi$ given in \Eq{Bolzeq}. To perform the numerical simulations, we set the following assumptions to simplify the setup: 

\noindent {\bf Flat Universe:}

\noindent We neglect the expansion of the Universe, i.e., $a(t) = 1$. Then, the second term on the left-hand side of \Eq{Bolzeq} vanishes.

\noindent {\bf Hierarchical timescales:}

\noindent We assume that $\chi_1$ and $\chi_2$ are always in thermal equilibrium and $f_{\chi_1,\chi_2}$ are given by Bose-Einstein or Fermi-Dirac distributions, \begin{align} f_{\chi_{1,2}} = f_{\chi_{1,2}}^{\rm eq} \equiv \left(e^{E_{\chi_{1,2}}/T} \mp 1\right)^{-1}, \end{align} where the minus (plus) sign is taken if $\chi_{1,2}$ are bosons (fermions), and $T$ is the cosmic temperature. 
{This assumption makes it possible to ensure exact energy-momentum conservation, while $\phi$ momenta are discretized and on the lattice in the numerical simulation. On the other hand, it is assumed that the interaction of $\phi$ is dominantly given by \Eq{process}. }

\noindent {\bf Initial conditions:}

\noindent We set the initial condition of \begin{align} f_{\phi}(t_{i}=0, p_{\phi}) = 0,
\end{align} for any $p_{\phi}$. 
This initial condition together with the previous assumption implies what we study is a thermal production process of $\f$.

\noindent {\bf Relativistic plasma:} We focus on the parameter region $T \gg m_{\chi_{1,2}}$.\\

In describing the behaviors of the distribution functions, it is convenient to introduce two time scales, $\Delta t_{\rm ignition}$ and $\Delta t_{\rm th}$. The former corresponds to the time scale of the onset of the stimulated emission of $\phi$: the occupation number or $f_\f$ of a certain momentum mode reaches unity. This time scale will be measured numerically in the simulation in this section and will be derived analytically in \ref{sec:Results}.
The latter one $\Delta t_{\rm th}$ corresponds to the time scale, which is conventionally called the thermalization time scale. $\Delta t_{\rm th}$ is calculated by the inverse of half of the production rate of $\phi$, \begin{align} \laq{intphi} \D t_{\rm th}^{-1} &\equiv \frac{1}{2}\G_{\phi},\non \\ 
\Gamma_\f &\equiv 2\frac{\pi^2 g_{\chi_1} g_{\chi_2}}{g_\f T^3} \int \frac{d^3\vec{p}_{\chi_1}}{(2\pi)^3} \frac{d^3\vec{p}_{\chi_2}}{(2\pi)^3} \sigma v_{\rm rel} f_{\chi_1}(p_{\chi_1}) f_{\chi_2}(p_{\chi_2}). \end{align}
Here the factor of $2$ accounts for the 2$\f$ production for each reaction. At this time scale, the number density of the thermal distribution is reached.

In the freeze-in scenario, it is the thermalization timescale, $\Delta t_{\rm th}$, that should be parametrically smaller than the time scale of the expansion of the Universe to produce the dark matter, while the time scale $\Delta t_{\rm ignition}$ is not usually considered.

The numerical results are shown in the figures.\footnote{ We discretized the momenta of $p_{\phi}$ and solved the unintegrated Boltzmann equation on each lattice.
For solving the differential equations, we used the techniques given in \cite{Yamaguchi:2016oqz} for solving the complecated renormalization group equations semi-analytically. }  In Figs.~\ref{fig:fphiBoson} and \ref{fig:p3fphiBoson}, solutions of the Boltzmann equation are shown in the planes ($p_{\phi}/T$, $f_{\phi}$) and ($p_{\phi}/T$, $(p_{\phi}/T)^3 f_{\phi}$), respectively, focusing on the case that $\chi_{1,2}$ are bosons. We set the power of the Mandelstam variable $s$ in \Eq{crosssect} {$n=1, 0, -1$} in each panel. We take $m_{\chi_1} = m_{\chi_2} = 0.1 {T},~m_{\phi} = 1 \times 10^{-8} {T}$, and $g_{\chi_1} = g_{\chi_2}{=g_\phi} = 1$. For comparison, we also plot the distribution for the thermal equilibrium $f_{\phi}^{\rm eq} = 1/(\exp(\sqrt{p_\phi^2 + m_\phi^2}/T) - 1)$ by the black dotted lines. The contours denote the spectra at different time slices. We see that {\it with $n < 1$, a significant production of the small-momentum mode occurs in a much shorter time scale than $\Delta t_{\rm th}.$} This time scale is $\Delta t_{\rm ignition}$. 

We can also see that at time scales much longer than $\Delta t_{\rm th}$, the populated IR modes get erased due to thermalization. However, in the freeze-in scenario, this time scale should never occur in order not to overproduce dark matter.

In Figs.~\ref{fig:fphiFermion} and \ref{fig:p3fphiFermion}, we show similar plots to the Figs. \ref{fig:fphiBoson} and \ref{fig:p3fphiBoson}, respectively but we focus on the case that $\chi_{1,2}$ are fermion. 
Except for ineternal degree of freedom {$g_{\chi_1}=g_{\chi_2}=2$}, the same input values as Figs.~\ref{fig:fphiBoson} and \ref{fig:p3fphiBoson} are used.  
Similar phenomena happen when $n<1$. In addition, we checked that with $n\ll -1$ and with $n$ being half integers below $1$ the significant production of the IR modes is maintained in both fermion and boson cases.

Our numerical results clearly show that in particular cases the stimulated emission occurs in the $2\to 2$ process. Those cases include the reaction via the renormalizable interaction $n\leq 0$.

\section{Analytical understanding}
\label{sec:Results}

In this part, let us understand what is happening behind the numerical simulations in the previous section. Let us call the production of the large population of the low momentum modes of $\phi$, \textbf{burst production}. The burst production occurred in the case of {$n<1$}. As one can see from the figures, this type of production happens if there is a hierarchy between $\Delta t_{\rm ignition}$ and $\Delta t_{\rm th}$, numerically ${\Delta t_{\rm ignition}}/{\Delta t_{\rm th}}\lesssim {\cal O}(10^{-2})$. 

To explain the dynamics we separate the burst production of $\phi$ into three stages~\cite{Yin:2023jjj}, i.e., (A) ignition, (B) burst production, and (C) saturation. The corresponding regimes are also shown in the figures.

\subsection{Ignition}

In the first stage (A), which is depicted by the green shaded regions in the figures, the modes of a low momentum, which we denote $p_{\phi}^{\rm burst}$, start to populate, and then the distribution function $f[p_{\phi}^{\rm burst}]$ reaches unity at $t\sim \Delta t_{\rm ignition}$.
As seen from the figures, the value of $\Delta t_{\rm ignition}$ is slightly different depending on the power $n$, but the peak momentum, $p_{\phi}^{\rm burst}$, is commonly given by $p_{\phi}^{\rm burst}\sim m^2_{\chi_{1,2}}/T$, independent of $n$. Let us understand those behaviors by analytically deriving the relevant formulas.

\subsubsection*{Ignition from the Small Angle Scattering}

As we can see from the figures, the IR modes that are produced significantly have peak momenta smaller than the masses of $\chi_{1,2}$. This motivates us to consider the small angle scattering of $\chi_1$ and $\chi_2$ with momenta of $\sim T$, which potentially produces a small momentum mode of $\phi$.

To study the small angle scattering analytically, we neglect all particles' masses in the kinematics, and later, we will introduce an IR cutoff on the center-of-mass energy relevant to the mass of $m_{\chi_{1,2}}$.
In this limit, we have $\sigma v_{\rm rel}\propto \frac{E^{n-1}_{\chi_1}E^{n-1}_{\chi_2}}{m_{\chi_1}^{n}m_{\chi_2}^{n}}(1-
\cos[\theta_{{\chi_1 \chi_2}}]
)^n$ in a factorized form. Here $\theta_{\chi_1\chi_2}$ is the angle between the momenta of $\vec{p}_{\chi_1}$ and $\vec{p}_{\chi_2}$ in the cosmic frame. 
Then we obtain the differential thermal production rate of $\phi$,
\begin{align}
&\partial_{\log p_{\chi_1}}\partial_{\log p_{\chi_2}}\partial_{\cos \theta_{\chi_1\chi_2}}\Gamma_{\phi}\simeq \frac{2\pi^2 c_n g_\phi }{T^3}
 \(1-
\cos\theta_{{\chi_1 \chi_2}}\)^{n}\non\\
& \times 
  \frac{g_{\chi_1} p^3_{\chi_1}}{(2\pi)^2 } \frac{g_{\chi_2}p_{\chi_2}^3}{2\pi^2} \frac{2^n E^{n-1}_{\chi_1}E^{n-1}_{\chi_2}}{32\pi m_{\chi_1}^{n}m_{\chi_2}^{n}} f_{\chi_1}[p_{\chi_1}]f_{\chi_2}[p_{\chi_2}]\\
&
 \sim \frac{c_n g_\phi g_{\chi_1}g_{\chi_2}T^{2n+1}}{128\pi^3  m_{\chi_1}^{n}m_{\chi_2}^{n}} \theta^{2{n}}_{{\chi_1 \chi_2}} \quad \text{when}~ p_{\chi_1}\sim p_{\chi_2}\sim T.
\end{align}
In the last approximation, we used $f_{\chi_{1,2}}(p_{\chi_{1,2}}\sim T)\sim 1$, replaced $E_{\chi_{1,2}}$ and $p_{\chi_{1,2}}$ with $T$, and assumed $|\theta_{\chi_1\chi_2}|\ll1$.

This small angle scattering of \Eq{process} in the cosmic frame is Lorentz boosted compared with the center-of-mass frame by the Lorentz factor of 
\beq 
\gamma \sim 1/\theta_{\chi_1 \chi_2}. 
\eeq 
This is because the typical energy of $E_{\chi_{1,2}}\sim T$ is $1/\theta_{\chi_1 \chi_2} $ times the center-of-mass energy of $ E_{\rm CM}\sim \theta_{\chi_1 \chi_2} T$. 
The produced $\phi$ particles have a broad spectrum compared to the center-of-mass frame in the range 
\beq \laq{range}
E_{\phi}\sim [T \theta_{\chi_1 \chi_2}^2 -T]. 
\eeq 
In particular, the lower bound is obtained when $\phi$ is produced in the direction opposite to the boosted direction within an angle of $\theta_{\rm CM} \sim \theta_{\chi_1 \chi_2}$ in the center-of-mass frame. 
Then, $E_\phi \sim E_{\rm CM}/\gamma=\theta_{\chi_1 \chi_2}^2T.$ 
Given the assumption, the differential cross section in the center-of-mass frame does not depend on the angular of injected $\phi$ particles,\footnote{A non-trivial angular dependence of the $\f$ production in the center-of-mass frame could change the analytic formula of the ignition rate~\Eq{criterion}. 
} simple kinematics shows that the probability for each reaction to produce $\f$ 
scales $
\frac{\pi}{4\pi} \theta_{\rm CM}^2\sim E_\f/\g E_{\rm CM} \sim E_\f/T$. Here $\theta_{\rm CM}\ll 1$ is considered.  
This represents the fraction of the surface area for the typical angle of scattering in the center-of-mass frame.
Namely 
\begin{align}
&\partial_{\log E_\phi} 
\partial_{\log p_{\chi_1}}\partial_{\log p_{\chi_2}}\partial_{\log \theta_{\chi_1\chi_2}}\G_{\phi}|_{p_{\chi_{1,2}\sim T}}\non \\
&\sim \frac{E_\phi}{T} \times \frac{c_n  g_{\chi_1}g_{\chi_2}g_\f T^{2n+1}}{128\pi^3 m_{\chi_1}^{n}m_{\chi_2}^{n}}  \theta^{2{n}+2}_{{\chi_1 \chi_2}}.
\end{align}

On the other hand, the production rate of the occupation number for $p_{\phi}\sim E_\f$ mode by this small angle scattering is obtained by dividing the increasing rate of the number density  by the phase space volume
\begin{align}
\laq{SA}
\dot{f}^{\rm SA}_\f(p_\phi)&\sim \frac{T^3/\pi^2}{{4\pi p_\phi^3/(2\pi)^3}} \partial_{\log E_\phi} 
\partial_{\log p_{\chi_1}}\partial_{\log p_{\chi_2}}\partial_{\log \theta_{\chi_1\chi_2}}\G_{\phi}|_{p_{\chi_{1,2}\sim T}}\\ 
&\sim  
\frac{1}{E_\f^2} \times \frac{c_n  g_{\chi_1}g_{\chi_2}g_\f T^{2n+3}}{64\pi^3 m_{\chi_1}^{n}m_{\chi_2}^{n}}  \theta^{2{n}+2}_{{\chi_1 \chi_2}}.
\end{align}
Thus, the smaller the $E_\f$ the larger the $\dot{f}_\f^{\rm SA}[E_\f]$ with a fixed $\theta_{\chi_1\chi_2}.$ 
Namely, the fastest incresing mode is $E_\f \sim \theta_{\chi}^2 T$, which has
\beq 
\laq{fSA}
\dot{f}^{\rm SA}_\f(p_\phi\sim \theta_{\chi_1\chi_2}^2 T)\sim  
 \frac{c_n  g_{\chi_1}g_{\chi_2}g_\f T^{2n+1}}{64\pi^3  m_{\chi_1}^{n}m_{\chi_2}^{n}}  \theta^{2{n}-2}_{{\chi_1 \chi_2}}.
\eeq
As a consequence we have analytically derived the condition for the small angle scattering to be important:
\beq 
\boxed{{n<1}} \laq{condn}
\eeq 
in which case 
the smaller the $\theta_{\chi_1\chi_2}$ the larger the $\dot{f}^{\rm SA}_\f(p_\phi\sim \theta_{\chi_1\chi_2}^2 T)$. This is also the criterion that the ignition time scale is faster than the thermalization time scale as we will see. 
From now on, we focus on the case with $n<1$.

Our estimation so far by neglecting the mass of $\chi_1,\chi_2$ is only valid when 
$
E_{\rm CM}\gtrsim m_{\chi_1} +m_{\chi_2} 
$ which gives the lower bound of $\theta_{\chi_1 \chi_2}$
\beq
\theta_{\chi_1 \chi_2}> \theta_{\chi_1 \chi_2}^{\rm min}\equiv c \frac{m_{\chi_1} +m_{\chi_2}}{T}
\eeq
with $c$ being an $\O(1)$ coefficient. This also places an upper bound for \Eq{fSA}. This immediately leads to the momentum for the mode that defines the ignition timescale 
\beq
\boxed{ p_{\f}^{\rm burst}\sim   \frac{(m_{\chi_1} +m_{\chi_2})^2}{T}.}
\eeq
i.e. the lowest momentum, $p_\f\sim \theta_{\chi_{1}\chi_2}^2 T$, for the lowest $\theta_{\chi_1 \chi_2}=\theta_{\chi_1\chi_2}^{\rm min}.$ This is consistent with the numerical simulation. 

The ignition time scale can also be obtained from the inverse of $f_\f^{\rm SA}(p_\f^{\rm burst})$
\beq 
\laq{criterion}
\boxed{\D t_{\rm ignition}^{-1}= C 2^{2n-2}c^{2n-2} \frac{c_n  g_{\chi_1}g_{\chi_2}g_\f T^{3}}{64\pi^3   m_{\chi_{1,2}}^{2}}  
}
\eeq 
Here, we assumed $m_{\chi_1}\sim m_{\chi_2}$ and used $m_{\chi_{1,2}}$ to denote both, and we introduced a parameter $C\sim \O(1)$ to take into account of our rough estimation with neglecting various $\O(1)$ factors. 
Interestingly, the scaling on $T$ is independent of $n$, and is similar to the $1\to 2$ case \cite{Yin:2023jjj}. For instance, the analytic formula with $c\approx 1.5, C\approx 9.5$ fits the numerical result for the bosonic case well. We also checked the agreement by varying $m_{\chi_{1,2}}$ and $n<-1, n=-1/2,1/2, 3/2$ etc., in the numerical simulation. 

At the time scale of $\D t_{\rm ignition}$, the mode of momentum of $p_\f^{\rm burst}$ has the occupation number of unity, and the system moves to the next stage.  

Before discussing the next stage, let us mention the 
 condition for the ignition to happen. We note that the rate \eq{criterion} is usually faster than the thermalization rate if \Eq{condn} is satisfied. In fact, when the integral \eq{intphi} is UV dominant\footnote{We can easily check that when $n>-1(-2)$, the integral of $\log p_{\chi_{1,2}}$ is dominated at $p\sim T$ for $\chi_{1,2}$ being bosons (fermions).
}
this is trivial because $\Delta t_{\rm th}$ has the same scaling as \Eq{fSA} with $\theta_{\chi_1\chi_2}$ set to 1. Since $|\theta_{\chi_1\chi_2}|\ll1$, the ignition time scale is faster than the thermalization time scale when $n<1$. When the integral in \Eq{intphi} is IR dominant, on the other hand, on the dimensional ground $\D t_{\rm th}^{-1}\propto \frac{T^2}{m_{\chi_{1,2}}} \AND m_{\chi_{1,2}}$ for the case $\chi_{1,2}$ are bosons and fermions, respectively. The scaling of $T^2$ comes from the distribution function, $f_{\chi_{1,2}}\sim T/E_{\chi_{1,2}}$, at the low momentum. In both cases, the thermalization time scale is slower than the ignition time scale \Eq{criterion}.  This is also consistent with the numerical simulation.

\subsection{Burst}
In the second stage (B), which is shown by the red-shaded region in the figures, the modes around $p_{\phi}^{\rm burst}$ that were populated in the first stage further grow on the time scale of $\O(\Delta t_{\rm ignition})$. An exponential growth or stimulated emission occurs, similar to the operating principle of a laser. To understand this, we observe that $S \simeq f_{\chi_1}[p_{\chi_1}]f_{\chi_2}[p_{\chi_2}]
\left(1+f_{\phi}[p_{\phi}] \right)$ in the collision term for $\dot{f}_\phi[p_\phi]$ when $f_{\phi}[p_{\phi}]\gtrsim 1$. Thus, $\dot{f}^{\rm SA}[p_{\phi}^{\rm burst}]$ should be replaced by $\dot{f}^{\rm SA}[p_{\phi}^{\rm burst}](1+f_{\phi}[p_{\phi}^{\rm burst}])\sim \dot{f}^{\rm SA}[p_{\phi}^{\rm burst}]f_{\phi}[p_{\phi}^{\rm burst}] $, which can be solved to obtain
\beq
f_\phi(p_\phi^{\rm burst})\sim e^{t /\Delta t_{\rm burst}}.
\eeq
Here the coefficient satisfies $\D t_{\rm burst}\sim \D t_{\rm ignition}$ as $\D t^{-1}_{\rm ignition}\sim \dot{f}^{\rm SA}[p_{\phi}^{\rm burst}]$.  
This burst production of $\phi$ is mainly dominated by the process \beq 
\laq{burstprocess}
\chi_{1}(p_{\chi_1}\sim T)\chi_{2}(p_{\chi_2}\sim T) \to \phi(p_{\phi_1}\sim p_\f^{\rm burst}) \phi(p_{\phi_2}\sim 2T),\eeq so that not only the IR modes but also UV modes around $p_{\phi}\sim 2T$ are produced prominently. 
This can be clearly seen in Figs.~\ref{fig:p3fphiBoson} and \ref{fig:p3fphiFermion} with $n<1$.

\subsection{Saturation}
In the third stage (C), which is also depicted by the red-shaded regions in the figures, the burst production is terminated at around the time $t=t_{\rm saturation}= \O(10)\Delta t_{\rm ignition}$, independent of $n$.

This is because when sufficient $\phi$ production occurs, 
the back reaction
\beq \phi(p_{\phi_1}\sim p_\f^{\rm burst})\phi(p_{\phi_2}\sim 2T)\to \chi_{1}(p_{\chi_1}\sim T) \chi_{2}(p_{\chi_2}\sim T)\eeq becomes active and stops further stimulated emission. Namely, the system reaches a quasi-equilibrium. This is  similar to the steady state in the operation principle of the laser. 
Analytically, this quasi-equilibrium is represented by the condition $S\simeq 0$ for the reaction relevant to the $p_\phi^{\rm burst}$ mode. 
Noting $f_\phi(p_{\phi}^{\rm burst})\gg 1$, we get 
\beq 
f_\phi[p_\phi\sim 2T]\simeq \frac{1}{e^{p_{\phi}/T}-1},
\eeq 
from $S=0$. This implies 
$p_{\phi}^3  f_\phi[p_{\phi}]|_{p_\phi\sim 2T}\sim T^3$. 
Since each $\phi[p_{\phi}\sim 2T]$ production is associated with a $\phi[p_{\phi}^{\rm burst}]$ production, the $p_\phi^{\rm burst}$ modes produced via the burst production also have number density $\sim T^3$. As a result, we get 
\beq
(p^{\rm burst}_\phi)^3 f_\phi(p_\phi^{\rm burst})\sim (2T)^3 f_\phi(2T)\sim T^3.
\eeq
This is consistent with the numerical results, as can be seen from Figs.~\ref{fig:p3fphiBoson} and \ref{fig:p3fphiFermion}. 
In particular, by requiring $(p_\phi^{\rm burst})^3 e^{t_{\rm saturation}/\Delta t_{\rm ignition}}=T^3$, we derive $t_{\rm saturation}=\O(10)\Delta t_{\rm ignition}$, which aligns with the figures. Given the $n$-independent $p_\phi^{\rm burst}$, $t_{\rm saturation}/\Delta t_{\rm ignition}$ is also independent of $n$.

The intermediate momentum range is suppressed before the thermalization time scale. 
The production rate of the intermediate mode slightly depends on whether the $\chi_{1,2}$ are bosons or fermions. 
The fraction $\dot{f}_{\phi}[m_{\chi_{1}}/T]|_{\chi_{1}:{\rm Bosons}}/\dot{f}_{\phi}[m_{\chi_{1}}/T]|_{\chi_{1,2}:{\rm Fermions}}$
can be estimated to be $\sim4T/m_{\chi_1}$ from the contribution of $S$, where we neglect the back reaction for the intermediate mode and assumed that $f_{\phi}\ll 1$ and $m_{\chi_{1,2}}\ll T$. 
This explains the slight difference in the figures for the bosons and fermions.

The number density at the saturation stage, $n_\phi \equiv g_\phi \int \frac{d p_\phi}{2\pi^2} p_\phi^2 f_\phi(p_\phi) \sim T^3$, is obviously too large to produce dark matter much heavier than keV. Moreover, it is dominated by the contribution of the energetic modes around $p_\phi \sim T$. 
This quasi-equilibrium is maintained until the thermalization time scale, at which point the system approaches conventional thermal equilibrium. 
In the freeze-in scenario, in which the thermalization is never reached, once the burst production occurs the peculiar dark matter distribution is likely to remain as in the saturation stage.

Since the burst production stage ends and the saturation stage starts at the same time scale, $\O(1-10)\Delta t_{\rm ignition}$, it is very clear that as soon as the ignition stage (A) finishes, the usual freeze-in estimation fails. Thus we must evade the iginition to start in freeze-in scenario, which will set a novel bound to the scenario in \Sec{limit}. 
On the other hand, the burst production can produce the eV dark matter coldly by slighly modifying the setup, which will be studied in \Sec{EVDM}.

\subsection{Justification of Neglecting the Other Scattering Processes} 
\lac{scat}
Before ending this section, let us comment on the impacts of the other scattering reactions, such as $\f(p_{\f_1}) \chi_1(p_{\chi_1}) \to \f(p_{\f_2}) \chi_2(p_{\chi_2})$, that we did not take into account for the sake of generality and simplicity. Given that this reaction has a rate slower than or comparable to the thermalization rate, it rarely dissipates the burst-produced modes into higher momentum modes before thermalization. 

However, depending on the detail of the reaction,\footnote{One example may be the case that the reaction is enhanced when the momentum exchange is small. A similar phenomenon may be found in Ref.\,~ \cite{Kurkela:2011ti} where the authors study the thermalization in weakly coupled pure Yang-Mills theory.}   there may be another stimulated emissions altering the low momentum distribution once the previously analyzed burst production happens because the smaller momenta has smaller phase space volume and the ignition timescale can be even faster. 
Since the reaction does not change the total number density, we expect that this reaction would contribute to giving a `thermal' distribution of $\f$ with an effective temperature of $T_{\rm eff}\sim p_\f^{\rm burst}\ll T.$ To guarantee the number conservation, a condensate around the zero modes of $\phi$ must be formed. 
The thermal equilibrium is maintained with a chemical potential $\m_\f$ for  $\f$ with $\m_\f=m_\f$ (see Ref.\,\cite{Batell:2024hzo} for a numerical simulation for a similar setup).
In this case, our conclusions for the caveat on the freeze-in estimate and for the motivation to the eV dark matter do not change since the total number density of $\f$ is still $\sim T^3$, and it may be even colder.
A further study to confirm this case with a concrete model will be conducted elsewhere.

\section{A Novel Limit to the Freeze-in Mechanism}
\lac{limit}
So far, we have discussed the stimulated emission of dark matter pairs from the thermal plamsa via the reaction \eq{process} by neglecting the expansion of Universe. 
In the expanding Universe, we need to take into account the redshift of momentum. The redshift rate is characterized by the Hubble parameter, $H$, i.e., the expansion rate. Our result can easily be applied to the case when the Hubble expansion is much slower than the typical time scale (see \cite{Yin:2023jjj} for the numerical check in the $1\to 2$ decay case). In freeze-in production, we assume \beq 
\D t_{\rm th}^{-1}\lesssim H \laq{condDM}
\eeq throughout the cosmological history. Here $\lesssim$ implies that it is parametrically small to produce non-negligible dark matter number density. 

We find that, in order to recover the usual freeze-in production estimate, we need to ensure that the ignition time scale, estimated in \Eq{criterion}, satisfies 
\beq 
\D t^{-1}_{\rm ignition}\ll H
\eeq 
at any period of the thermal history if there is no dilution of the burst-produced modes. 
Indeed, for $n<1$
\beq 
 \D t^{-1}_{\rm th} \ll \D t^{-1}_{\rm ignition}.
\eeq 
It is non-trivial for having the condition \Eq{condDM} to be satisfied to produce correct abundance of the dark matter. 

In the following, we derive the bound to have a conventional freeze-in production during the radiation-dominated Universe depending on the nature of $\chi_{1,2}$ masses.

\subsection{Case of Vacuum Mass}
Let us consider the case where the vacuum mass of $\chi_{1,2}$ is not negligible, which is the case we mainly studied previously. Then $\D t_{\rm ignition}^{-1} \propto a^{-3}$ if $T\propto a^{-1}$. Here we remind that $a$ is the scale factor. In the radiation-dominated Universe, $H=T^2/M_{\rm pl} \sqrt{g_\star(T) \pi^2/90}\propto a^{-2}$, and thus one can find that $\D t_{\rm ignition}^{-1}/H$ is largest at the beginning of the radiation-dominated Universe, i.e., at the end of the reheating phase. Here $g_{\star}(T)$ is the relativistic degrees of freedom for the energy density, and $M_{\rm pl}\approx 2.4\times 10^{18}\GEV$ is the reduced Planck scale.
What we need is to check if $\D t_{\rm ignition}^{-1}/H\lesssim 1$ is satisfied at the end of the reheating.\footnote{One can also easily see that during the reheating phase, the ratio is IR dominated because of the diluting plasma.}

This gives a constraint for the reheating temperature, i.e., the highest temperature in the radiation dominated Universe,
\beq
\boxed{T_R\lesssim \frac{\pi^4 \sqrt{g_\star}   2^{\frac{15}{2}-2 n} c^{2-2 n}}{3 \sqrt{5} c_n C g_{\chi_1}g_{\chi_2} g_\f }\frac{m_{\chi_{1,2}}^2}{M_{\rm pl}}.}
\eeq

Asumming the late time dominance of the freeze-in production, which should be the case $n< 1/2$, we can estimate the dark matter abundacne from $\Omega_{\rm \f}^{\rm FI}\sim \frac{s_0}{\rho_c} m_{\f}\left.\frac{\Gamma_\f}{H} \frac{T^3}{\pi^2 s}\right|_{T=m_{\chi_{1,2}}}.$ Namely we consider that the dark matter is mostly produced at $T=m_{\chi_{1,2}}$.
Here $s=T^3 g_{\star,s}2\pi^2/45 $ is the entropy density, $s_0$ the present value, and $\rho_c$ the present critical density. 
Setting $\G_{\f }\sim c_n \frac{T^{2n+1}}{32 \pi^3 m_{\chi_{1,2}}^{2n}},$\footnote{This is justisfied when $n>-1(-2)$ for $\chi_{1,2}$ being bosons (fermions). } and using the requreiment $\Omega_{\rm \f}^{\rm FI}\sim 0.3$  we can replace $c_n$ by the $\phi$ mass
\beq 
T_R \lesssim 100\GEV \times 9^{-n}\frac{m_{\chi_{1,2}}}{\GEV} \frac{m_{\f}}{10\KEV}
\eeq 
where we assumed $g_\f=g_{\chi_{1,2}}=1, g_{\star}=g_{\star,s}=106.75$ 
for simplicity and used the fit value $c=1.5,C=9.5$.

Assuming the early time dominance scenairo, which corresponds to $1/2<n$  the dark matter is mostly produced around the end reheating. Thus $\Omega_{\rm \f}^{\rm FI}\sim \frac{s_0}{\rho_c} m_{\f}\left.\frac{\Gamma_\f}{H} \frac{T^3}{\pi^2}\right|_{T=T_R}.$ 
Similary, we can obtain the bound. For instance with $n=3/4$ we have $T_R\lesssim 3\GEV \frac{m_{\chi_{1,2}}}{1\GEV}\(\frac{m_{\f}}{10\KEV}\)^2.$ 
Note that with $n\geq 1$ the burst production does not occur and we do not have the constraint. We see that the bound restricts the reheating temperature not to high compared to the mass scale of the thermal plasma.

\subsection{Case of Thermal Mass: Higgs Portal-like Interaction}
When the vacuum mass is very small, we need to take into account the thermal reaction. Even the reaction \Eq{process} can induce a thermal mass for the particles. Thus, the contribution is inevitable when the vacuum mass is very small.   
To study this possibility concretely, we consider $\chi_{1,2}=\chi$, a scalar boson coupled to the dark matter via a portal interaction
\beq 
{\cal L}\supset -\l_P |\chi|^2\frac{\phi^2}{2}.
\eeq 
The amplitude of the reaction (with the symmetry factor) can be estimated 
\beq 
|{\cal M}|^2=c_0= \frac{1}{2}\l^2_P 
\eeq 
which has $n=0$.

We consider the mass squared for $\chi$ in the form of 
 \beq
 m_{\chi}^2= m_{\chi,0}^2+ A T^2
 \eeq
  with $A$ being the model-dependent parameter. For the Standard Model-like Higgs boson, we have $A\sim y_t^2/4$ (see, e.g.,Ref.\,~\cite{Carrington:1991hz}).

When the thermal mass contribution is dominant, $\D t_{\rm ignition}^{-1} \propto T\propto a^{-1}$. In the radiation-dominated Universe, $\D t_{\rm ignition}^{-1} /H $ is larger at lower temperatures. On the other hand, when the vacuum mass dominates, the ratio is larger at higher temperatures, as before. Therefore, we need to check that $\D t_{\rm ignition}^{-1} /H<1$ when $T=T_{\rm crit}$ satisfying $m_{\chi,0}^2\sim  A T_{\rm crit}^2$, assuming $m_{\chi,0}^2>0$.\footnote{In the case of negative $m_{\chi,0}^2$, the condition does not change much for the $2\to 2$ reaction, although a phase transition/crossover happens. In this case, we need to also include the $1\to 2$ process to produce the dark matter. A thermal mass of the mother particle in the $1\to 2$ process can be studied similarly by replacing the vacuum mass with the thermal one in \cite{Yin:2023jjj}. } 
 Then we obtain
 \begin{align}
&\boxed{A\gtrsim \frac{45 C^2 g_\chi^4 g_\phi^2 \lambda_P^4 M_{\rm pl}^2}{2097152 \pi^8 c^4 g_{\star}(T_{\rm crit})  m_{\chi,0}^2}}\\
&~~~ \approx 0.2 g_\chi^4g_\f^{2} \(\frac{\l_P}{10^{-6}}\)^4\(\frac{100\GEV}{m_{\chi,0}}\)^2
 \end{align}
 where in the last approximation, we again used the fit $c=1.5,C=9.5$ for the ignition time scale, and $g_\star=106.75$.
One should note that this is a very stringent bound requiring $A$, which is, roughly speaking, the squared of the coupling, to be larger than $\O(0.1)$ for the sample parameters we chose.
Here, $\l_P$ is chosen to yield the correct dark matter abundance from the $2\to 2$ freeze-in for $m_\f\sim \KEV$. One can see that this requires the thermal mass of $\chi$ to be large enough. 
For the Standard Model-like Higgs boson, fortunately, we have $A= \O(1)$ due to the sizable top Yukawa coupling, $y_t$, and the dominant production of the lighter dark matter is
 via the decay~\cite{Lebedev:2021xey}, leading to a smaller $\l_P$ for the correct dark matter abundance from the freeze-in. Thus, our constraint is not important. However, for generic portal dark matter, e.g., with an extended Higgs sector, the limit should be seriously considered.

\section{Some related topics}
\lac{app}
\subsection{Dark Matter Mass Effect on Burst Production}

\begin{figure*}[htbp!]
\includegraphics[scale=0.5]{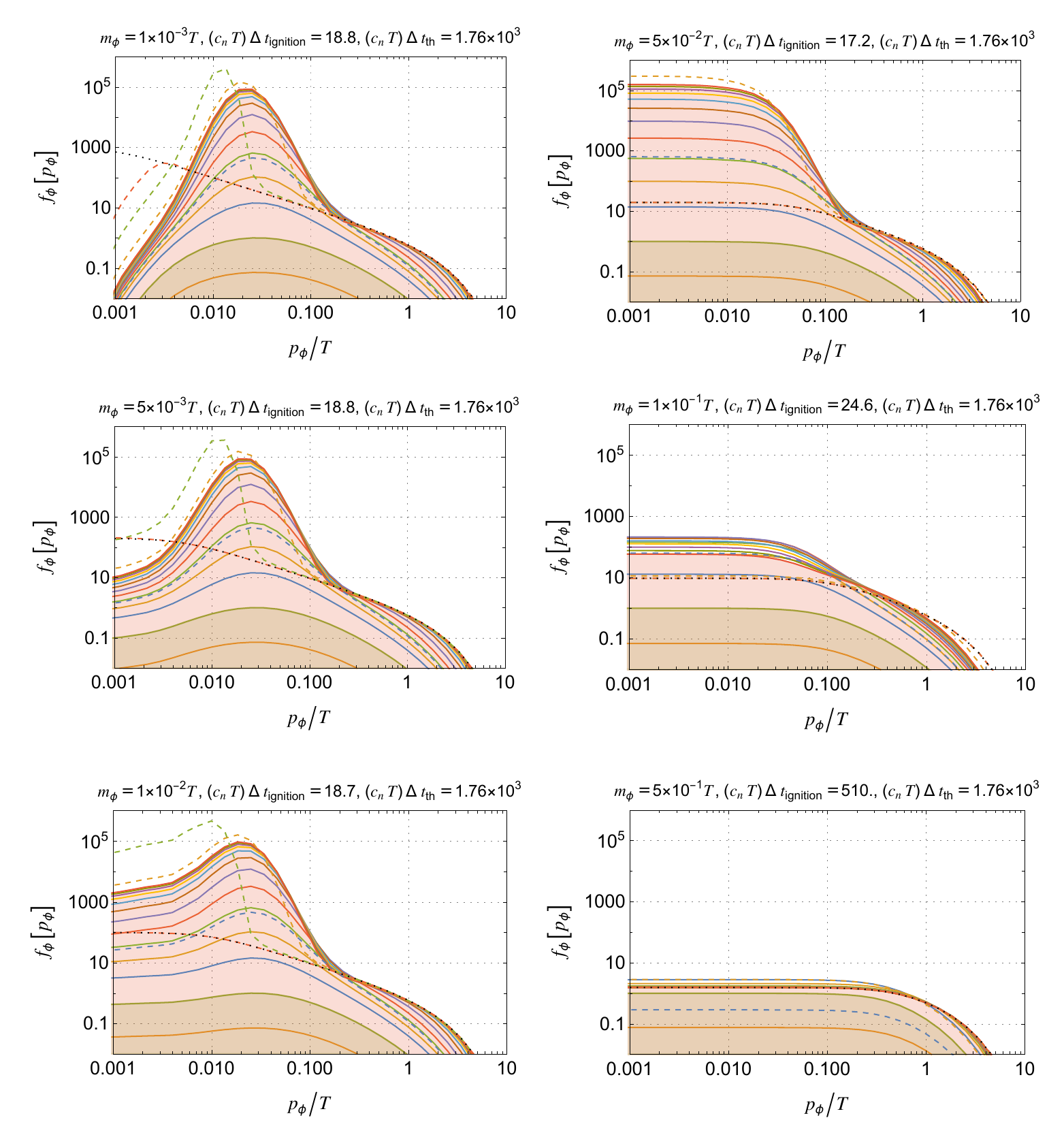}
\caption{The distribution function $f_{\phi}$ with non-negligible $m_\f$. we focus on the case where $n=0$ and $\chi_{1,2}$ are bosons. 
The mass of $\phi$ is set by $m_\phi/T=\{10^{-3},5\times 10^{-3},10^{-2},5\times 10^{-2},10^{-1},5\times 10^{-1}\}$ for each panel. Other inputs are taken to be the same as in Fig.~\ref{fig:fphiBoson}. 
}
\label{fig:mphidep}
\end{figure*}

\begin{figure*}[htbp!]
\includegraphics[scale=0.5]{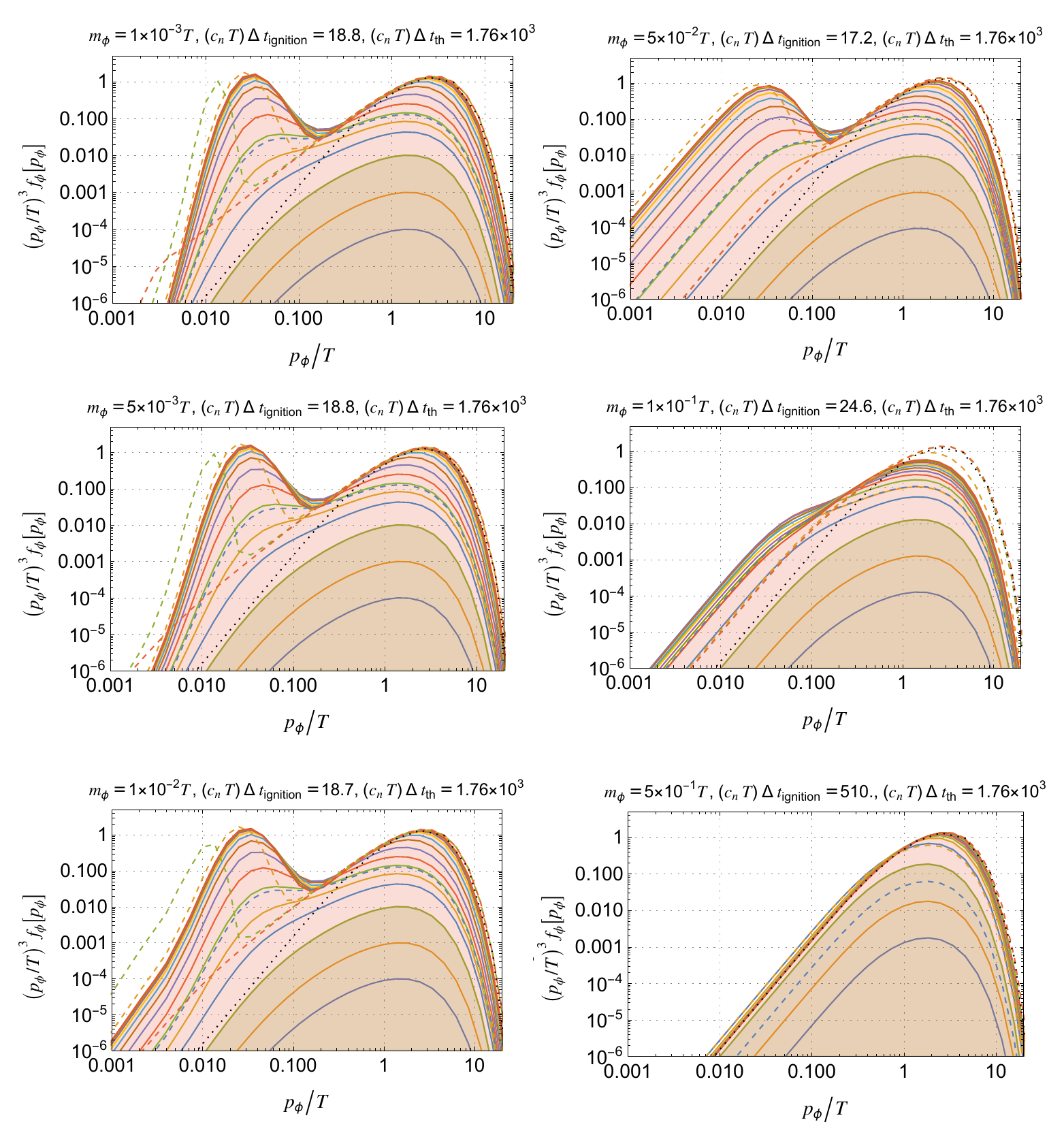}
\caption{The same plots with Fig.~\ref{fig:mphidep} but for $(p_{\phi}/T)^3f_{\phi}$. 
}
\label{fig:mphidep_p3f}
\end{figure*}

So far, we have neglected $m_\phi$ in discussing the production of dark matter analytically and taking the negligibly small mass in the simulation. Here, let us comment on the case where $m_\phi$ is not very small. An enhancement of the stimulated emission was analytically shown in the two-body decay of a scalar condensate~\cite{Moroi:2020has}, and in lattice simulations of a similar setup~\cite{Adshead:2023qiw} (see also \cite{Moroi:2020bkq} for the relation between the Boltzmann equation and narrow resonance). Indeed, such an enhancement also exists when we consider the two-body decays $\chi_1\to \chi_2 \phi$ when the masses of $\chi_1$ and $\chi_2$ are close while $m_\phi$ is small~\cite{Yin:2023jjj}. These effects occur because the phase space volume is more suppressed, and the Bose enhancement becomes more significant.

However, in this setup, we find that the mass efect does not enhance the iginition time scale as well as the number density of the burst producion of $\phi$ pair, as shown in Figs.~\ref{fig:mphidep} and \ref{fig:mphidep_p3f}. 
However, when the mass is parametrically heavy, $m_\phi \sim m_{\chi_{1,2}}^2/T$, the far IR mode is populated. 

In particular, an interesting behavior is that we have an almost constant $f_\phi$ below the mass. 
This can be understood from kinematics. Let us consider producing $\phi$ with momentum much lower than $m_\phi$, $p_\phi\ll m_\phi$, from the small angle scattering. Then in the cosmic frame~\footnote{The same argument works in the $1\to 2$ process by replacing $E_{\rm CM}$ by the mass of the mother particle.}
\begin{align}
\laq{Ef}
E_\phi &= \gamma \frac{E_{\rm CM}}{2} - \gamma \beta \cos \theta_{\rm CM} p_{\phi,\,\rm CM}, \\
\laq{ppara}
p_\phi^{\parallel} &= -\gamma p_{\phi,\,\rm CM} \cos \theta_{\rm CM} + \gamma \beta  \frac{E_{\rm CM}}{2},\\
\laq{pperp}
p_\phi^{\perp} &=  p_{\phi,\,\rm CM} \sin \theta_{\rm CM}.
\end{align}
Here we have to take into account the mass effect, and $p_{\phi,\,\rm CM}$ denotes the momentum of $\phi$ in the center-of-mass frame:
\beq 
p_{\phi,\,\rm CM} = \sqrt{E_{\rm CM}^2/4 - m_\phi^2}.
\eeq 
From \Eq{Ef}, we see that $\frac{d E_\phi}{d \cos \theta_{\rm CM}} = \gamma \beta p_{\phi,\,\rm CM}$, and we can derive
\beq 
\frac{d \cos\theta_{\rm CM}}{d p_\phi}= \frac{p_\phi}{E_\phi \gamma \beta p_{\phi,\,\rm CM}}.
\eeq 
As we mentioned, the distribution of $\phi$ production in the center-of-mass frame is spherically symmetric, and we have a flat distribution of the $\phi$ production rate with respect to $\cos \theta_{\rm CM}$ for any reaction. This means that producing $p_\phi \ll E_\phi\simeq m_\phi$ is suppressed by $p_\phi$. 

In addition, we also find from \Eq{Ef} with $E_\phi\sim m_\phi$
\beq 
\gamma \sim \frac{E_{\rm CM}}{m_\phi}.
\eeq 

Furthermore, from \Eq{pperp}, we have 
\beq 
\theta_{\rm CM}<  \frac{p_\phi}{p_{\phi,\,\rm CM}},
\eeq 
which implies that the phase space integral is suppressed by $p_\phi$. As a result, we have a $p_\phi^2$ suppression in the production rate of the $p_\phi$ mode. To get the occupation number (or distribution function), we need to divide it by the phase space volume $\propto p_\phi^2 dp_\phi$, and therefore, the growth of the $p_\phi (\ll m_\phi)$ mode is independent of $p_\phi$. Since the energy has $E_{\rm CM} \sim T/\gamma$ for the burst production, which is irrelevant to $p_\phi$, the ignition rate and the momentum modes for the backreaction for any $p_\phi\ll m_\phi$ has the almost same momentum modes of particles that $\f(p_{\f}\ll m_\f)$ interacts with, resulting in $p_\phi$-independent growth and saturation. 

\subsection{eV Dark Matter}
\lac{EVDM}

\begin{figure*}[htbp!]
\includegraphics[scale=0.5]{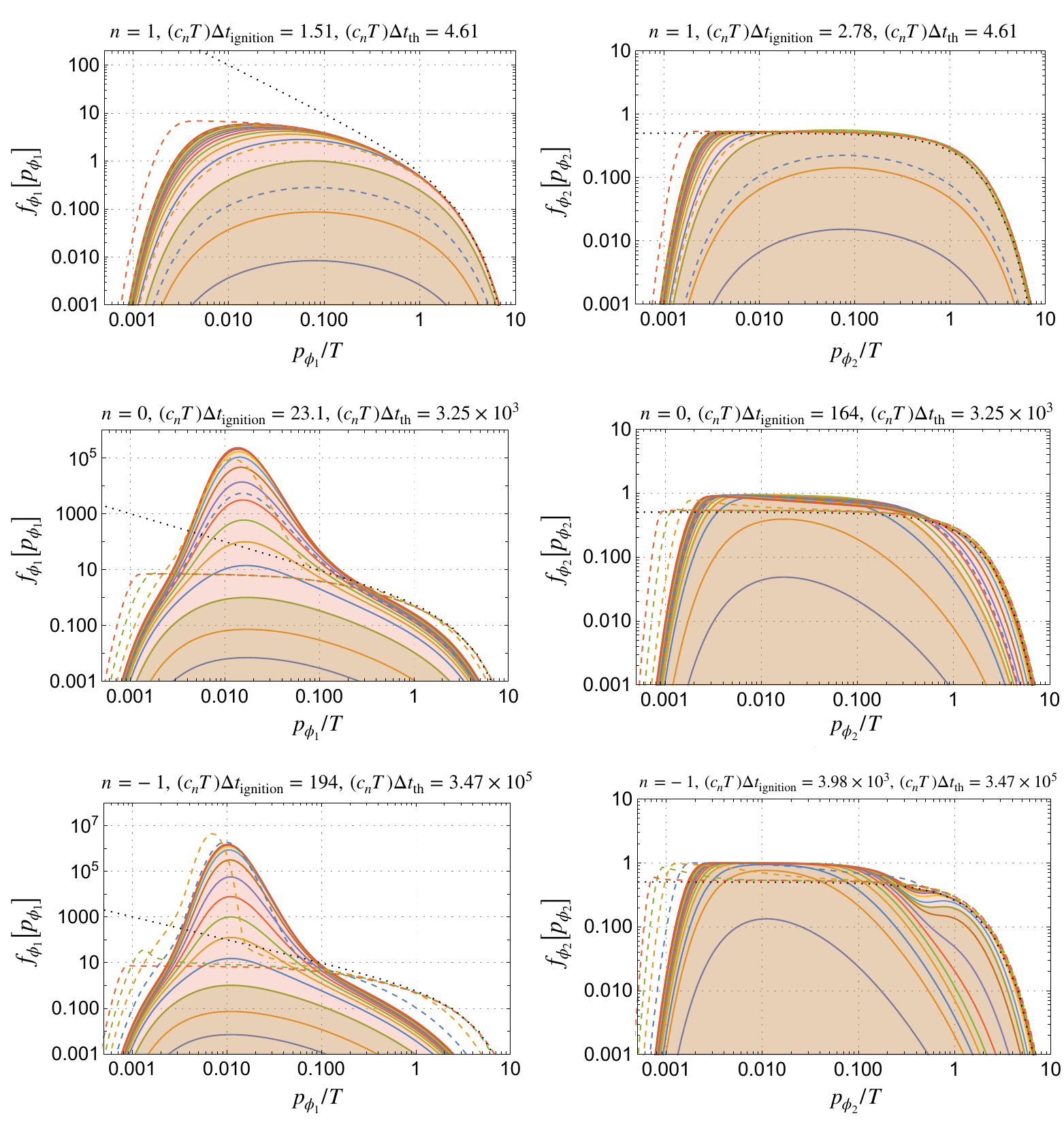}
\caption{The distribution functions $f_{\phi_1}$ and $f_{\phi_2}$ for the process $\chi_{1}(boson) \chi_{2}(fermion)\to \phi_1(boson) \phi_2(fermion)$  in the left panels and right panels, respectively.
}
\label{fig:FBFB_f}
\end{figure*}

\begin{figure*}[htbp!]
\includegraphics[scale=0.5]{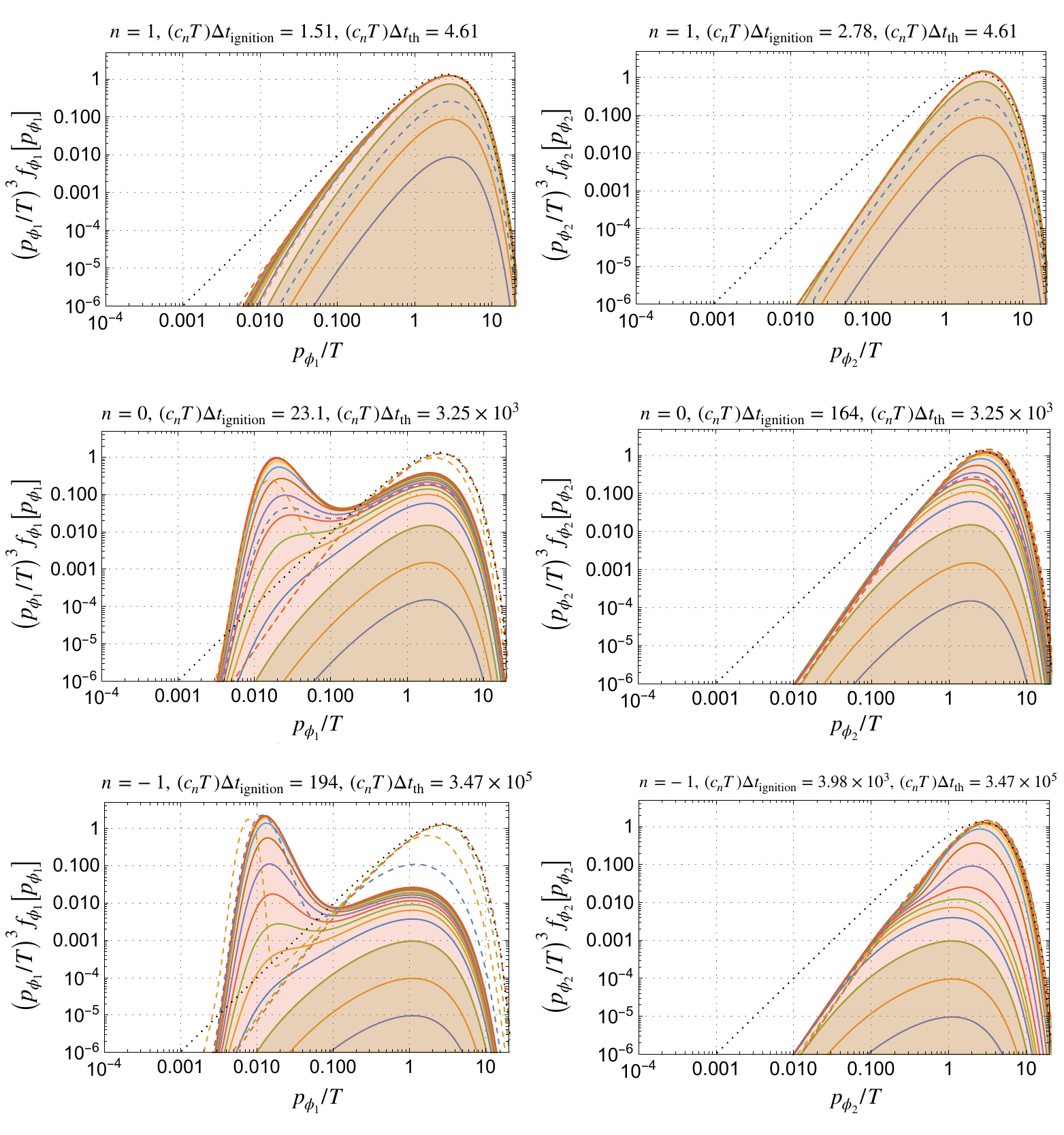}
\caption{The same plots as Fig.~\ref{fig:FBFB_f} but for $(p_\f/T)^3f_\f$.
}
\label{fig:FBFB_p3f}
\end{figure*}

So far, we have discussed avoiding the burst production of $\phi$. Here, let us consider a cosmological scenario where burst production occurs resulting in the cold dark matter in the eV mass range. The $p_\phi^{\rm burst}$ mode can be a successful cold dark matter candidate with the eV mass range because the number density is $\sim T^3$ and the short free-streaming length~\cite{Yin:2023jjj}. On the other hand, the $p_\phi\sim 2T$ component in \Eq{burstprocess} also has a $\sim T^3$ number density, contributing to the too much of hot dark matter. Thus, the high momentum mode needs to be somehow suppressed to achieve a consistent cosmology with the eV cold dark matter. 

One way to suppress the hot component is to consider some new interactions providing a cooling effect for the high-energy modes. Alternatively, one can consider a similar reaction 
$\chi_1\chi_2\to\phi_1 \phi_2$,
with $\phi_2$ being a fermion, or massive scalar, while $\f_1$ is the bosonic dark matter.\footnote{Alternatively, a light scalar with large degrees of freedom, $g_{\f_2}\gg 1$ may also work. In this case, the ignition of $\f_2$ other than the dark matter, $\f$, is suppressed by $g_{\f_2}$.} Note that eV dark matter is light and the stability is easily guranteed without imposing a symmetry. 
In this case, the burst production of $\phi_2$ at the low-energy mode, which corresponds to the high momentum mode of $\f_1(p_{\f_1}\sim 2T)$, is absent. Then $\phi_1$ at high momentum mode is not produced, but $p_{\f_1}\sim p_{\f}^{\rm burst}$ mode is produced as well. This is actually shown in Figs.~\ref{fig:FBFB_f} and \ref{fig:FBFB_p3f}.

\section{Conclusions}\label{sec:Discussion}

In this paper, we have investigated the stimulated emission effect for light boson production via the two-to-two scattering processes, i.e., Eq.~\eqref{eq:process}, assuming the initial states are thermalized. In our analysis, we have numerically solved the so-called unintegrated Boltzmann equation without neglecting Bose enhancement/Pauli blocking factors in the flat Universe.  To apply to generic models, we considered the amplitude squared scaling with  $s^{n}$.

We found that with $n<1$, the IR modes ($p_{\phi}\sim m_{\chi_{1,2}}^2/T$) of a boson $\phi$ are violently produced on a time scale much shorter than the usual thermalization time scale. 
 This is due to small angle scattering of the mother particles, which can produce bosonic dark matter with suppressed momentum, concentrated in a small phase space volume. If the threshold, the occupation number to be unity, reaches,  stimulated emission occurs. Depending on the mother particle's (thermal) mass, studies of freeze-in production should account for this effect. In addition, we found that eV-range dark matter can be produced coldly due to this effect. The eV mass arises from the known temperature of matter-radiation equality and the fact that the backreaction of the stimulated emission leads to a steady state where the dark matter number density is approximately equal to that of the mother particles, i.e., the thermal plasma. Thus, the dark matter in the eV mass range is theoretically well-motivated and special, similar to the hot dark matter paradigm.

\acknowledgments
This work was supported by JSPS KAKENHI Grant No. 20H05851 (W.Y.), 21K20364 (W.Y.), 22K14029 (W.Y.), and 22H01215 (W.Y.), 23KJ0086 (K.S.), the National Science Centre, Poland, under research Grant No. 2020/38/E/ST2/00243 (K.S.), and Incentive Research Fund for Young Researchers from Tokyo Metropolitan University (W.Y.).

\bibliography{reference}
\end{document}